\documentclass{scspaperproc}
\usepackage{latexsym}
\usepackage{graphicx}
\usepackage{mathptmx}

\usepackage{url}
\usepackage{pdfpages}
\usepackage{amsmath}
\usepackage{amsfonts}
\usepackage{amssymb}
\usepackage{amsbsy}
\usepackage{amsthm}

\usepackage[colorinlistoftodos,prependcaption,textsize=small]{todonotes}
\usepackage{blindtext}
\presetkeys%
    {todonotes}%
    {inline,backgroundcolor=yellow}{}

\newtheoremstyle{scs}
{3pt}
{3pt}
{}
{}
{\bf}
{}
{.5em}
{}

\theoremstyle{scs}

\usepackage{graphicx}

\usepackage{amssymb}
\usepackage{lineno}
\usepackage{booktabs}
\usepackage{float}
\usepackage{subcaption}

\usepackage[pdftex,colorlinks=true,urlcolor=blue,citecolor=black,anchorcolor=black,linkcolor=black,bookmarks=false]{hyperref}

\begin{document}

\pagestyle{fancyplain}

\thispagestyle{plain}
\firstPageHead{}

\chead{\fancyplain{}{\itshape\small Serena, Zichichi, D'Angelo, and Ferretti \vspace{8pt}}}

\rhead{}
\cfoot{}
\renewcommand{\headrulewidth}{0pt} 

\makeatletter
\let\@internalcite\cite
\def\cite{\def\@citeseppen{-1000}%
    \def\@cite##1##2{(##1\if@tempswa , ##2\fi)}%
    \def\citeauthoryear##1##2##3{##1 ##3}\@internalcite}
\def\citeNP{\def\@citeseppen{-1000}%
    \def\@cite##1##2{##1\if@tempswa , ##2\fi}%
    \def\citeauthoryear##1##2##3{##1 ##3}\@internalcite}
\def\citeN{\def\@citeseppen{-1000}%
    \def\@cite##1##2{##1\if@tempswa, ##2)\else{}\fi}%
    \def\citeauthoryear##1##2##3{##1 (##3)}\@citedata}
\def\citeA{\def\@citeseppen{-1000}%
    \def\@cite##1##2{(##1\if@tempswa , ##2\fi)}%
    \def\citeauthoryear##1##2##3{##1}\@internalcite}
\def\citeANP{\def\@citeseppen{-1000}%
    \def\@cite##1##2{##1\if@tempswa , ##2\fi}%
    \def\citeauthoryear##1##2##3{##1}\@internalcite}
\def\shortcite{\def\@citeseppen{-1000}%
    \def\@cite##1##2{(##1\if@tempswa , ##2\fi)}%
    \def\citeauthoryear##1##2##3{##2 ##3}\@internalcite}
\def\shortciteNP{\def\@citeseppen{-1000}%
    \def\@cite##1##2{##1\if@tempswa , ##2\fi}%
    \def\citeauthoryear##1##2##3{##2 ##3}\@internalcite}
\def\shortciteN{\def\@citeseppen{-1000}%
    \def\@cite##1##2{##1\if@tempswa, ##2\else{}\fi}%
    \def\citeauthoryear##1##2##3{##2 (##3)}\@citedata}
\def\shortciteA{\def\@citeseppen{-1000}%
    \def\@cite##1##2{(##1\if@tempswa , ##2\fi)}%
    \def\citeauthoryear##1##2##3{##2}\@internalcite}
\def\shortciteANP{\def\@citeseppen{-1000}%
    \def\@cite##1##2{##1\if@tempswa , ##2\fi}%
    \def\citeauthoryear##1##2##3{##2}\@internalcite}
\def\citeyear{\def\@citeseppen{-1000}%
    \def\@cite##1##2{(##1\if@tempswa , ##2\fi)}%
    \def\citeauthoryear##1##2##3{##3}\@citedata}
\def\citeyearNP{\def\@citeseppen{-1000}%
    \def\@cite##1##2{##1\if@tempswa , ##2\fi}%
    \def\citeauthoryear##1##2##3{##3}\@citedata}
%
%
%
\def\@citedata{%
    \@ifnextchar [{\@tempswatrue\@citedatax}%
                  {\@tempswafalse\@citedatax[]}%
}

\def\@citedatax[#1]#2{%
\if@filesw\immediate\write\@auxout{\string\citation{#2}}\fi%
  \def\@citea{}\@cite{\@for\@citeb:=#2\do%
    {\@citea\def\@citea{, }\@ifundefined
       {b@\@citeb}{{\bf ?}%
       \@warning{Citation `\@citeb' on page \thepage \space undefined}}%
{\csname b@\@citeb\endcsname}}}{#1}}%

%
\def\@citex[#1]#2{%
\if@filesw\immediate\write\@auxout{\string\citation{#2}}\fi%
  \def\@citea{}\@cite{\@for\@citeb:=#2\do%
    {\@citea\def\@citea{, }\@ifundefined
       {b@\@citeb}{{\bf ?}%
       \@warning{Citation `\@citeb' on page \thepage \space undefined}}%
{\csname b@\@citeb\endcsname}}}{#1}}%

%
\def\@biblabel#1{}
\makeatother

\newdimen\bibindent
\bibindent=.25in

\def\thebibliography#1{\section*{\refname}\list
   {}{\settowidth\labelwidth{[#1]}
   \leftmargin \bibindent
   \itemindent -\bibindent
   \listparindent \itemindent
	 \itemsep 4pt
   \parsep 0pt
   \usecounter{enumi}}
   \def\newblock{}
   \sloppy
   \sfcode`\.=1000\relax}

\setlength{\baselineskip}{12.7pt}

\includepdf[pages={1}]{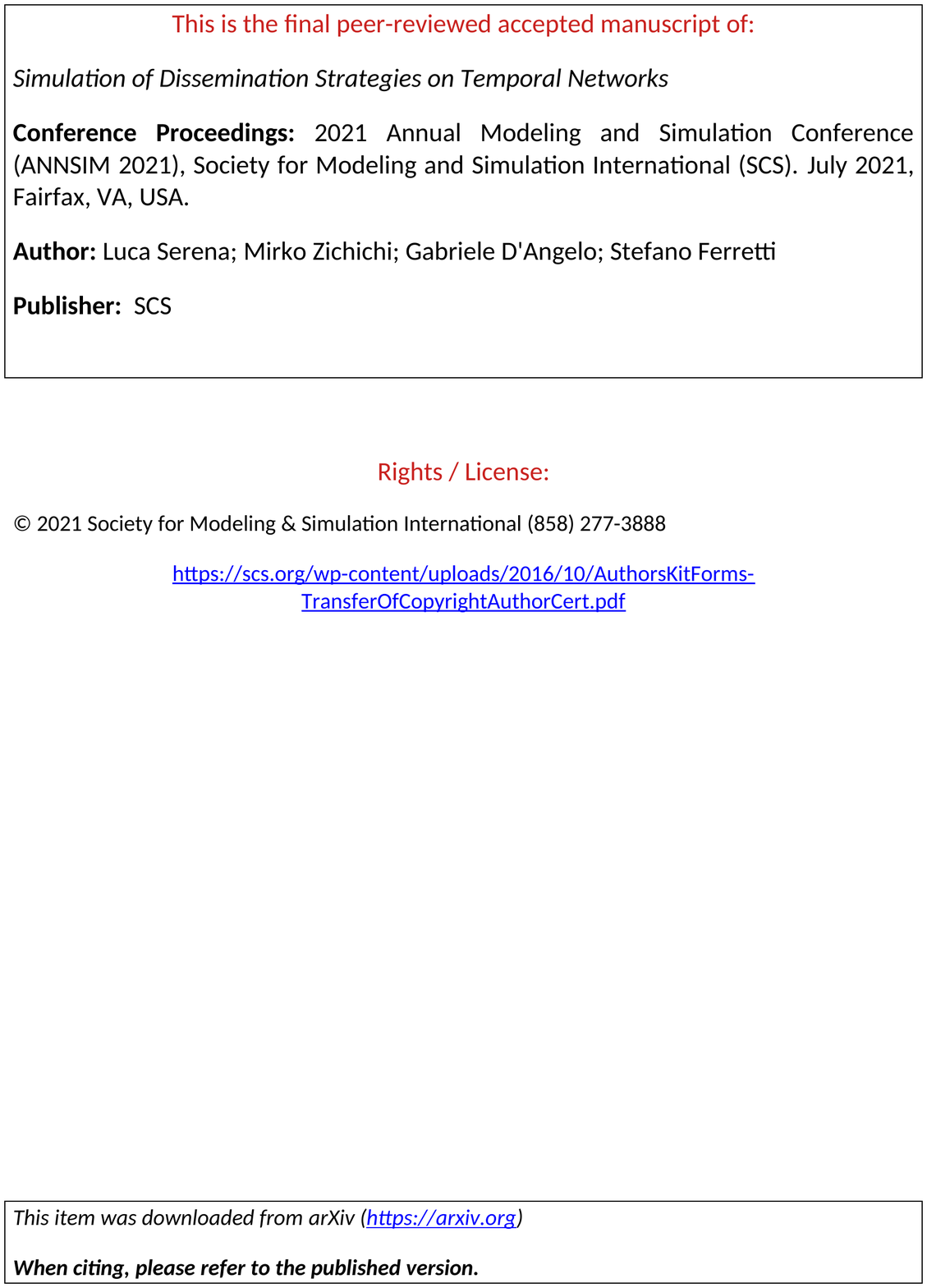}

\def\SCSconferenceacro{ANNSIM'21}
\def\SCSpublicationyear{2021}
\def\SCSconferencedates{July 19-22}
\def\SCSconferencevenue{Fairfax, VA, USA}

\title{Simulation of Dissemination Strategies on Temporal Networks}

\author{
Luca Serena\\[12pt]
CIRI ICT, University of Bologna\\
Via dell'Università, 336 - Cesena (FC), ITALY
\\luca.serena2@unibo.it
\and
Mirko Zichichi\\[12pt]
OEG, Universidad Politécnica de Madrid\\ 
ETSIINF, Boadilla del Monte (MD), SPAIN
\\ mirko.zichichi@upm.es.
\and
Gabriele D'Angelo\\[12pt]
DISI, University of Bologna\\
Mura Anteo Zamboni, 7 - Bologna, ITALY
\\g.dangelo@unibo.it
\and
Stefano Ferretti\\[12pt]
DiSPeA, University of Urbino ``Carlo Bo''\\
Piazza della Repubblica, 13 - Urbino, ITALY
\\stefano.ferretti@uniurb.it
}

\maketitle
\begin{abstract}
In distributed environments, such as distributed ledgers technologies and other peer-to-peer architectures, communication represents a crucial topic. The ability to efficiently disseminate contents is strongly influenced by the type of system architecture, the protocol used to spread such contents over the network and the actual dynamicity of the communication links (i.e.~static vs. temporal nets). In particular, the dissemination strategies either focus on achieving an optimal coverage, minimizing the network traffic or providing assurances on anonymity (that is a fundamental requirement of many cryptocurrencies). In this work, the behaviour of multiple dissemination protocols is discussed and studied through simulation. The performance evaluation has been carried out on temporal networks with the help of LUNES-temporal, a discrete event simulator that allows to test algorithms running on a distributed environment. The experiments show that some gossip protocols allow to either save a considerable number of messages or to provide better anonymity guarantees, at the cost of a little lower coverage achieved and/or a little increase of the delivery time.
\end{abstract}
\section{Introduction}
Nowadays, the ubiquitous cloud computing paradigm implies that most of the applications running on the Internet follow a centralized client-server approach. This means that all the resources of the application are situated in some servers, and the users of the system need to contact such servers in order to retrieve the information. An alternative to this scheme is to use a decentralized approach, where the data and computation resources are distributed among the various nodes and the central servers, if present, only have a coordination role. So, it is possible to design systems whose architecture is decentralized and where all the nodes, often referred as peers, share the workload without privileges, a hierarchy and central entities being involved, i.e. Peer-to-Peer (P2P).\\
\indent Normally, P2P systems make use of an overlay network, meaning that an application level communication network is created, running on top of an already existing network infrastructure (i.e.~the Internet), often regardless of the real geographical distribution of involved nodes. In such a case, this scheme might lead to a lot of traffic overhead to keep the network up and running~\shortcite{p2ptypes}. However, it has been demonstrated that it is possible to considerably reduce the network traffic by using a smart approach to propagate the information on the network~\shortcite{ddf}. Often there has been little interest in traffic minimization in distributed environments, because it may not be crucial for the functioning of the system. Thus, usually peers relay the new data that they receive to all their neighbors (except the one from which they received the message), and the only concern is to avoid infinite loops of messages. However, for certain applications, traffic minimization can be a relevant issue, and significant improvements can be achieved without compromising the efficiency of the communication. Several algorithms to spread the messages among the peers exist and, depending on the features of the system, certain protocols (and certain protocols' parameters) may turn out to be more appropriate than others.\\
\indent Simulation is a useful methodology in order to investigate which protocols are more suitable for the various purposes. By generating a virtual environment where multiple nodes communicate through the use of messages, it is possible to analyze the behaviour of the different algorithms and to evaluate the overall efficiency with the help of some performance metrics. Generally, it is desirable to achieve a very high if not complete coverage (i.e.~the percentage of peers that receive a message), and to minimize the network traffic and the delivery time (i.e.~the time between the creation of a message and its delivery). However, no algorithm can maximize all these features simultaneously, so it is necessary to find a balanced trade off, taking into account which trait is more important for the specific application. For example, blockchains usually do not require all the nodes to suddenly receive all the blocks and transactions, because some information, if missing, can be retrieved. Therefore, in this scenario a protocol that focuses on anonymity and traffic minimization might be convenient, but without totally neglecting the delivery time, that if too large could lead the network to reach an inconsistent state.
The output of the metrics can be influenced by many factors other than the gossip protocol, like the connectivity of the network or the topology of the overlay. Another factor to consider is the dynamicity of the nodes, which are not necessarily always active, but they can turn on and off over time, therefore changing the structure of the graph. Models associated with such kinds of dynamic systems are usually referred to as temporal networks, which are the main focus of this work.\\
\indent In this paper we introduce LUNES-temporal, a simulator for distributed environments, focusing on two types of dissemination algorithms: protocols for enhancing the anonymity (that is a feature considered as necessary in most modern cryptocurrencies and related blockchains~\shortcite{khalilov2018survey}) and protocols for minimizing the network traffic (i.e.~improve the communication efficiency). Temporal networks have been used to carry out our investigation in a dynamic environment, where nodes are randomly located and they can change their activation state.\\
\indent From our results, it turns out that some dissemination algorithms are more efficient than others, either because they ensure both good anonymity properties and 100\% of messages delivery (i.e.~Dandelion++\shortcite{dandelion++}) or because they allow to save network traffic under the same coverage achieved. Furthermore, it was noticed that the presence of hubs (i.e.~nodes with a number of links that greatly exceeds the average) in the graph (used to represent the network) is relevant for which concern the average delivery time.\\
\indent The remainder of the paper is organized as follows. Section 2 introduces some background and related work. Section 3 describes the design choices of the software tool and deals with the critical aspects of the implementation. Section 4 analyzes the results obtained by testing different dissemination protocols in various network configurations. Finally, Sections 5 provides some concluding remarks.

\section{Background and Related Work}
In this section, we introduce some background and related work needed by the rest of the paper. In particular, we briefly describe Distributed Ledger Technologies (DLTs), their underlying P2P networks used to manage communication in DLTs and the main issues related to their simulation.

\subsection{Peer-to-Peer Systems as Temporal Networks}
Some years ago, the P2P technology got attention for making it possible to share and exchange resources such as text files, music, video. Furthermore, for some time it appeared to be the main component of decentralized Internet (e.g.~ BitTorrent, Emule, Gnutella and TooFast). Although interest in this technology was fading due to the growth of cloud computing, a new wave of interest emerged  with the advent of blockchains and cryptocurrencies. Blockchains, and DLTs in general, are systems in which a ledger of transactions is distributed among several nodes. Different types of DLTs provide different implementations of the ledger, but in blockchains, transactions are collected in blocks and each block contains the hash of the previous one. To achieve decentralized verification of each block, the entire blockchain is replicated among all nodes forming a P2P network. These nodes operate in an unstructured topology where each peer is randomly connected to other peers and transactions are disseminated across the entire network. Each node then independently verifies the transactions received in order to ensure their consistency and to avoid double spending attacks \shortcite{nakamoto2009bitcoin}. 
Once blocks are created they are disseminated in the network as well, to notify the peers that the blockchain has been extended.\\
\indent Generally speaking, P2P applications usually run on top of an already existing network, like the Internet, and the underlying overlay network can be represented as a graph (directed or undirected depending on the implementation), where the nodes are the peers of the system and the edges connect the nodes that are directly in touch. Different P2P architectures may exist, usually depending on the type of the application that must be supported.
Two prominent aspects related to the functioning of a P2P system are i) how the overlay is built and maintained in order to cope with churns (i.e.~nodes that dynamically come and go in the system) , ii) how messages are exchanged among peers.\\
\indent In a P2P environment, a node is not connected with all the other peers of the network: that would be possible just for applications with a very small number of actors involved. As a result, each node is connected only with a certain number of other nodes, and thus  multiple message relays among peers are needed in order to spread the information to the whole network. For example, in Bitcoin each node has a number of connections in the range [$8$, $125$]~\shortcite{btcflood}. How these connections are chosen is fundamental, since they drive the creation of overlays with specific topologies. Moreover, the dynamicity of peers that can (freely) join and leave the network, often requires the use of some protocol to maintain the net healthy and connected~\shortcite{ddf,Ferretti2013481}.\\
\indent The modeling of this system dynamicity has been a main topic of research in the past~\shortcite{SURATI2017705}. Studies were focused on the evolution of the overlay, and usually did not consider applications running on top of it. On the other hand, studies on (the modeling of) distributed applications typically assume that the underlying overlay is static, thus neglecting the possible presence of churns. In this paper, we deal with both these issues by means of a specific simulator of temporal networks. More specifically, through temporal networks, we are able to model the dynamism of interactions among nodes. On top of this, the simulator allows running specific message dissemination models.\\
\indent As concerns how messages are exchanged among nodes in the P2P overlay, several dissemination schemes can be devised, they are discussed in detail in the next subsection.

\subsection{Dissemination Protocols}\label{sec:dissemination}
A gossip protocol is a dissemination scheme that aims to spread a message within a P2P network. In our work, we will consider the following algorithms~\shortcite{d2009simulation}:

\begin{itemize}
\item \emph{Probabilistic Broadcast} (PB). Given a forwarding parameter $p$, there is $p\%$ of chance that a node forwards the message to all the neighbors except the forwarder and $(100-p)\%$ that it does not send it to any other node.
\item \emph{Fixed Probability}(FP). Given a parameter $p$, a roll of dice for each of the neighbors (except the forwarder) takes place, so each neighbor will have $p\%$ of chance of receiving the message.
\item \emph{Fixed Fanout}(F-FAN). Given a parameter $n$, the message is forwarded to $n$ neighbors. If the concerned node has less than $n$ neighbors, then it is forwarded to all the neighbors.
\item \emph{Degree Dependent Function Algorithms}(DDF). The message is sent based on the degree (i.e.~number of neighbors of a node) of the potential receiver: the less connected is the neighbor, the more possibilities it has to receive the message~\shortcite{ddf}. The information of the degree should lead to a better propagation, since the protocol aims to reach the most isolated nodes of the graph, which are the most likely to cause problems in the dissemination process. Above all in networks with a high degree differential among the nodes, this strategy could lead to better results, at the cost of sending messages from time to time for updating the data about neighbors' connections. In our implementation, if the potential receiver has less than $3$ neighbors, then the message is always forwarded. Otherwise, a strategy based on either logarithmic or exponential functions is followed. Therefore, given a parameter $X$, the message is forwarded with either probability $1/log_D X$ or $1/D^X$, where $D$ represents the degree of the potential receiver.
\item \emph{Dandelion}. The focus of this protocol is to prevent de-anonymization attacks~\shortcite{amarasinghe2019survey}. It consists of two phases: initially in the so-called ``stem phase'' a message is relayed to just one neighbor chosen at random. Then, after a certain number of stem steps, the messages are forwarded to all the neighbors (i.e.~``fluff phase'').
\item \emph{Dandelion++}. This algorithm has been proposed to improve the Dandelion protocol, as some assumptions are not necessarily valid in practice and the theoretical properties of anonymity and guaranteed delivery may not always be ensured. The protocol is currently used by cryptocurrencies like Monero~\shortcite{monero} and Zcoin~\shortcite{dissemination}. Among the changes from the previous version there is a different policy for transactions forwarding. Time is divided into epochs and in each epoch a peer is either a ``diffuser'' (spreading messages to all the neighbors) or a ``relayer'' (sending messages to just another node), depending on the hash of the node’s own identity and on the epoch number. Furthermore, a fail-safe mechanism is implemented, preventing data to get lost if a malicious or a defective node did not correctly relay the disseminated message.
\end{itemize}

In all the considered dissemination strategies except for Dandelion and its variations, when a new message is originated then it is sent by its creator to all his neighbors. This is done in order to avoid getting stuck immediately at the first step.
Independently from the protocol being used, a cache system and a Time-To-Live (TTL) are employed to prevent infinite loops of messages. The strategy is respectively to not forward an already received message and to allow a message to be relayed only for a limited number of hops.\\
\indent The metrics used to evaluate the performance of such algorithms are the following:

\begin{itemize}
\item \emph{Successful communication rate}. It indicates the percentage of times that a node is able to get in touch with a designated node.
\item \emph{Messages sent}. It indicates the average number of messages being sent in the process of getting in touch with the recipient node. Both successful and unsuccessful attempts are considered in the count.
\item \emph{Delay}. It indicates the number of discrete steps needed on average to contact the recipient node. Except protocols with recovery mechanisms, it corresponds to the number of hops needed for the delivery.
\item \emph{Time overhead}. It indicates the ratio between the delay obtained with a certain gossip protocol and the delay obtained with a pure broadcast scheme, which guarantees (by definition) to obtain the lowest delay that can be possibly achieved.
\end{itemize}

Certain applications may need to achieve a 100\% successful communication rate in order to function properly, but in other cases a minor percentage could still be acceptable to make everything work fine. An example of the second case is the blockchain, where most of the messages being sent are blocks or transactions. Transactions are validated by different nodes, meaning that it is not indispensable for all the actors involved in the validation process to know exactly all the transactions that have been spread into the network. For what concerns the blocks, specific recovery mechanisms are used by the blockchain nodes to retrieve non-delivered blocks when it is detected that there are missing blocks. Anyway, although 100\% of successful communication rate is not indispensable for certain systems, often a very high percentage is still required for a correct functioning.

\section{LUNES-temporal} 
LUNES (Large Unstructured Network Simulator) is a time-stepped discrete event simulator for complex networks, which permits to simulate certain network protocols and to evaluate their efficiency. The software (available at \url{ https://pads.cs.unibo.it}) is thought to work with distributed environments, which can be customized by choosing topology, size and level of connectivity of the network~\shortcite{sf-gda}. LUNES is implemented on top of ART\`IS/GAIA simulation middleware, which offers the primitives for time management and communication among simulated entities, enabling also a parallel and distributed execution~\shortcite{ddf}. The tool is thought to be easily expandable, so that adding new simulation models on top of the existing services is a rather simple task. For example, the software was also used to simulate some popular security attacks to different setups of blockchains.\\
\indent LUNES-temporal is a specific version of LUNES in which the simulator is able to dynamically manage the overlay network used to simulate the message dissemination. It allows to model a large number of peers, each one is labeled with an integer identifier and a state indicating whether the node is active or not. Active nodes act normally by receiving and forwarding messages. On the contrary, the non-active nodes are deactivated, meaning that they are temporarily off and they cannot receive and relay messages until a future reactivation. When a node deactivates, it immediately removes all its outgoing connections. At the next timestep, its former neighbors will be notified of the deactivation and in turn they will remove such a peer from its neighbors list. 
The nodes, if active, can directly communicate with a set of neighbors, with the messages needing one timestep to be delivered to the adjacent recipient.\\
\indent In the original version of the software (i.e.~LUNES) the network is modelled as a static graph, meaning that it is not subject to changes over time. This approach is useful when it is necessary to deal with precise configurations (i.e.~topology, number of nodes/links and network diameter). However, even if static networks can still be a valid tool for the modelling of many systems, in some cases they lack veracity. In our specific case, peer-to-peer applications usually work in dynamic environments, where continuously new nodes become members of the system and other nodes temporarily (or permanently) shut down.\\
\indent The employment of a temporal network, as the core of the simulator, permits to work with a dynamic environment, whose structure changes over time. At the beginning, a certain number of connections (i.e.~links) between nodes are established, but the network configuration mutates more or less rapidly over time depending on the activation rate of the temporal model. This leads to a complete renovation of the structure of the graph after a certain amount of timesteps. It may happen that, due to the random transformations, certain nodes achieve a much bigger degree than other ones, but such conditions are fleeting. For most connected networks random changes are not a big concern, since the protocols still efficiently succeed in the dissemination, but for few connected networks the risk of bottleneck is real. In a graph where some nodes have very few connections the possibilities that certain messages get lost are greater, above all for probabilistic gossip algorithms that do not consider the degree of the neighbors. Therefore, a strategy to control and manage the evolution of a temporal network (for example by setting a minimum number of active connections for each peer) might play an important role for dissemination scopes.\\
\indent In LUNES-temporal, the overlying layer implements the application layer message exchange protocol. In particular, this module implements all the types of dissemination algorithms described in Section~\ref{sec:dissemination}. The dissemination is performed in accordance with the rules of the chosen gossip protocol, which can be selected by the user before the simulation is started.

\section{Results and Analyses}

\subsection{Setup and Methodologies}
The P2P environment used in this performance evaluation consists of {10\,000} peers, including both active (the majority) and deactivated nodes.  In our model, the graph is undirected, meaning that the link relationship benefits from the commutative property. The only case where a link is available only in one direction is when a node has still to be notified about the deactivation of a neighbor.\\
\indent At each discrete step of simulation, there is a specific probability of changing the activation state for a node. The chance for an active node to deactivate, and the chance for a deactivated node to activate, are not equal. Otherwise, the graph would tend to have 50\% of active nodes over time. Instead, such probabilities are tuned to maintain approximately the same number of active peers over time. Specifically, in the following simulations we have, on average, 80\% of active nodes, with a probability of 1\% for a deactivated node to activate and a probability of 0.25\% for an active node to shut down. When a node activates or when a node remains without neighbors, it connects with a certain number of active peers, chosen at random (i.e.~with uniform distribution) along the graph. There is no further criterion for the nodes attachment, so the graph retains its random properties.\\
\indent The experiments that we executed consist of over {10\,000} epochs (i.e.~portion of the simulation where a single communication attempt between peers is performed) in which data are collected to calculate the metrics described in Section~\ref{sec:dissemination}. At the beginning of each epoch an applicant and a holder of a hypothetical resource are chosen at random among the active nodes of the system and these two peers will remain active for all the duration of the epoch. If the applicant manages to get in touch with the holder then the communication has success, otherwise a fail occurs. The average number of messages sent is obtained by taking into account the data of all the epochs while, of course, in order to calculate the average delay just the successful epochs are considered. 
Each message that is created has a Time-To-Live (TTL) that sets the duration of its epoch. Usually, a TTL of $15$-$20$ timesteps is abundantly sufficient, but the algorithms with a fail-safe mechanism, above all under certain configurations, may require more time for the delivery. So in these cases it is necessary to increase the duration of the epochs.\\
\indent Some specific simulations have been run to observe the behaviour of the network over time and it turned out that the graph does not split into multiple components, moreover it results that the diameter of the graph does not change significantly. A split of the graph into multiple components would entail the inability to deliver certain messages, and the growth of the graph's diameter would likely increase the average delay. Before testing the performance of the dissemination protocols, some experiments have been made in order to evaluate how the average degree of the nodes and the level of dynamism of the network can influence the results. The obtained outcomes are reported in the following.

\subsection{Average Degree Variation}
\begin{figure}[h]
    \centering
    \begin{subfigure}[b]{0.495\textwidth}        \includegraphics[width=\textwidth]{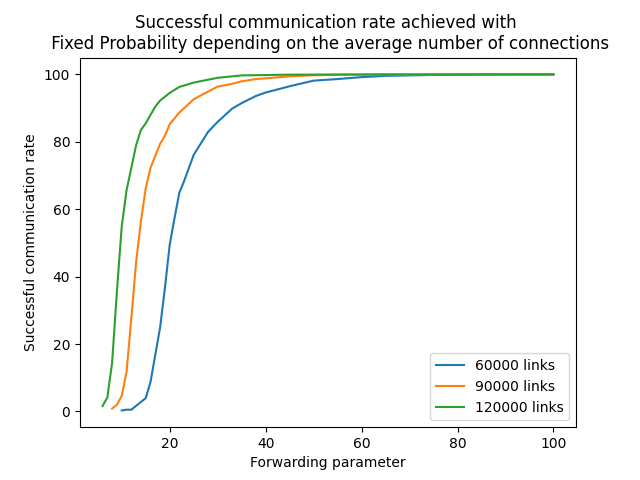}
    \end{subfigure}
    \begin{subfigure}[b]{0.495\textwidth}
    \includegraphics[width=\textwidth]{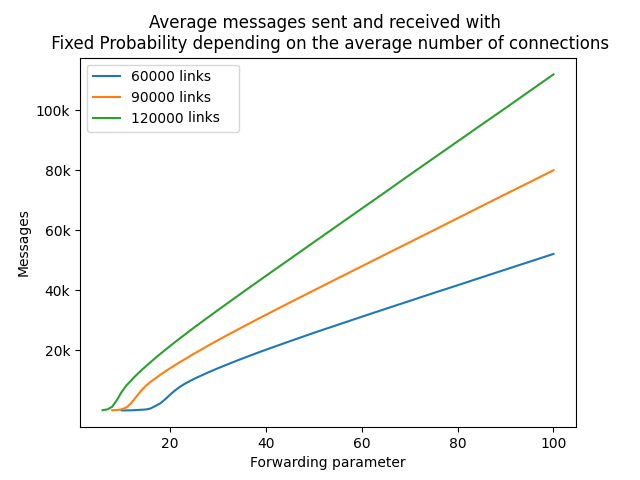}
    \end{subfigure}
    \begin{subfigure}[b]{0.495\textwidth}   
    \includegraphics[width=\textwidth]{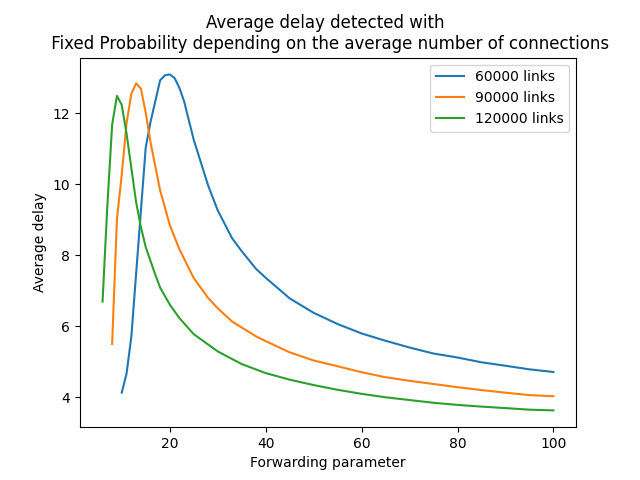}
    \end{subfigure}
    \caption{Effects of the average number of edges per node on specific performance metrics.}
    \label{connections-change}
\end{figure}

Figure~\ref{connections-change} represents how the output of the metrics changes depending on the average number of connections per node. As expected, under the same protocol configuration, the most-connected networks achieve a higher successful communication rate and a lower average delay, but at the cost of a greater volume of forwarded messages. Therefore, in case one is interested in minimizing the network traffic, it is necessary to calibrate the parameters of the dissemination protocols according to the average number of connections.

\subsection{Dynamism Rate Variation}
Specific experiments have been carried out to test how the level of network dynamism can influence the dissemination. The main effect caused by the variation of the graph structure is that sometimes a message is lost. In fact, when a node $A$ sends at time $t$ a message to the neighbor $B$, it can happen that $B$ deactivates right at the timestep $t$, so it will not be able to receive the message at the timestep $t+1$ as expected. Only at $t+1$ the node $A$ will be notified of the deactivation of $B$ and then it will proceed to remove the link. 
In our default configuration the chance for a node to shut down is 0.25\%, meaning that on average a message out of $400$ will be lost in this way. Raising this percentage is equivalent, in terms of dissemination, to reduce the probability of forwarding a message. So, for example, using Fixed Probability in a static network with $n$ as forwarding parameter allows to achieve the same successful communication rate when using $n-x$ in a network where at each timestep on average $x\%$ of the node deactivates. However, it is necessary to consider that it is not realistic to represent a network with a very high dynamism rate. In fact, in our modelling a single timestep represents the time necessary to receive, check and possibly forward a message, so it is unrealistic to have a big percentage of the nodes to deactivate in such a small amount of time. The only dissemination algorithm for which these variations of the graph's configuration are crucial is Dandelion, since messages can be lost in the stem phase. This issue, though, can be easily fixed by employing a fail-safe mechanism. So, in conclusion, we can state that when having a small and reasonable dynamism rate of the network, variations of such a value do not have a significant influence on the performance of the gossip protocols.

\subsection{Dandelion vs. Dandelion++}
\begin{figure}[t]
  \centering
   \begin{subfigure}[b]{0.495\textwidth}   
    \includegraphics[width=1\textwidth]{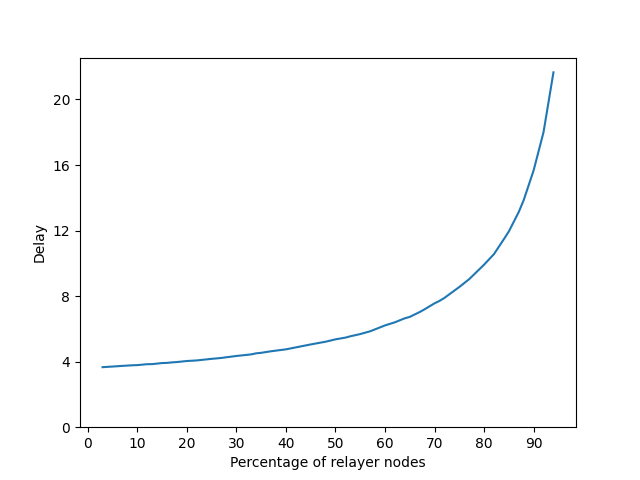}
  \caption{Dandelion++, average delay variation depending on the percentage of relayer nodes.}
  \label{dandelion++-delay}
\end{subfigure}
   \begin{subfigure}[b]{0.495\textwidth}   

    \includegraphics[width=1\textwidth]{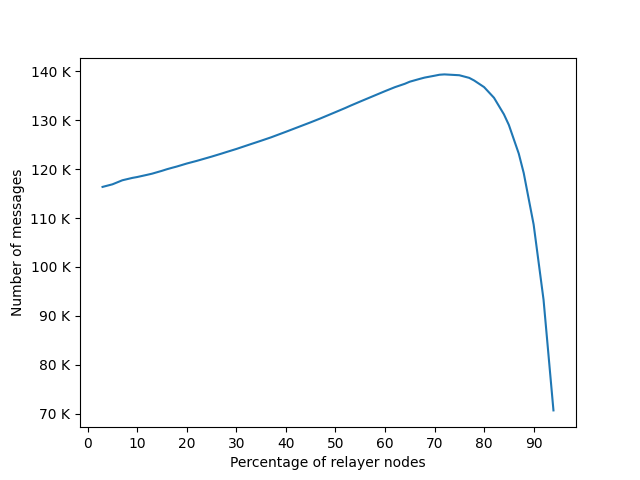}
  \caption{Dandelion++, average messages sent variation depending on the percentage of relayer nodes.}
  \label{dandelion++-messages}
  \end{subfigure}
  \caption{Dandelion++ metrics.}
\end{figure}

The main difference of Dandelion and Dandelion++ with respect to the other dissemination algorithms considered in this paper, is the goal to increase the anonymity properties in blockchain-like systems. In practice, this is achieved by making more difficult to link the transactions to the Internet address (i.e.~IP address) that generated them~\shortcite{dandelion++}. In fact, commonly used gossip protocols spread the information isotropically along the graph, thus making it possible for an observer peer of the system to understand where a certain transaction has originated, after having observed the spread of messages through the network over time.
Certain applications, like Bitcoin, introduce random independent delays during the dissemination in order to avoid timing attacks, but such a strategy is not considered sufficient to avoid deanonymization~\shortcite{bojja2017dandelion}. The ``ping pong'' mechanism implemented in the Dandelion stem phase makes it more difficult to pinpoint who originated the message, even by a peer connected to most or all the nodes or by attackers who know significant information about the overall topology of the network. On the other hand, Dandelion when used in a blockchain presents certain drawbacks: during the stem phase a node might not relay the message as according to the protocol, either because it acts maliciously or simply because it is off at the moment. Resilience against Sybil attacks (i.e.~attacks where several nodes purposely do not relay the messages of certain users, trying to isolate them from the network) was analyzed in a previous paper~\shortcite{dissemination}, pointing out that by integrating Dandelion with a fail-safe mechanism then a P2P system manages to remain operational even when facing a big amount of peers acting maliciously.
In this performance evaluation, even if we do not consider the presence of attacks performed by other nodes, Dandelion still has some problems to achieve a 100\% coverage. This is due to the fact that sometimes happens (during the stem phase) that a node shuts down right in the moment when it was supposed to receive and then relay a message. This issue, in the same way as for Sybil attacks, can be fixed by triggering the fail-safe mechanism when needed: a node is therefore able to start the fluff phase by itself, in the event that after relaying a message during the stem phase it did not receive it back in the fluff phase after a certain amount of time.\\
\indent Dandelion++, in order to avoid the messages from getting lost, implements the fail-safe mechanism and, in addition, has a different management for forwarding transactions while ensuring anonymity. In each epoch (about $10$ minutes) a node is either a ``diffuser'' or a ``relayer'', and therefore stem phase and fluff phase are not temporally distinct and separated. Some tests have been carried out changing each time the percentage of chance for a peer to be a relayer, with that decision depending on the node ID and on the epoch number. To get better guarantees against deanonymization, it is advised to have 80\% or more of relayer nodes.\\
\indent From our tests it turned out that, independently from the network topology, 100\% of successful communication rate can be easily achieved, as long as the network is sufficiently connected and the fail-safe mechanism is implemented.
From a coverage point of view, the protocols with the fail-safe mechanism behave like a pure broadcast scheme, because if a problem during the ``delicate'' one-relay phase occurs, then the broadcast diffusion will still take place. 
If one was also interested to the minimization of the messages being sent, it would be possible to integrate Dandelion or its variants with one of the previously mentioned dissemination algorithms: in this case Fixed Probability or Degree Dependent Function protocols would be applied in the fluff phase (or by diffuser nodes for Dandelion++), instead of simply spreading the messages to all the neighbors.
The recovery mechanism entails a greater number of forwarded messages and a slightly bigger average delay: this is due to the times when the fail-safe mechanism is triggered (and without which it would not be possible to deliver the message). In these situations, the message is delivered later, since a certain amount of time has to pass before the fluff phase is started. The delay gap, between $0.01$ and $0.25$, grows proportionally with the number of stem steps.\\
\indent Dandelion++, having a fail-safe mechanism implemented, guarantees 100\% of average successful communication rate as well, independently from the percentage of diffusers and relayers. However, as shown in Figure~\ref{dandelion++-delay}, the percentage of relayers has a big influence on the average delay, both because the dissemination process gets slower and because it happens more frequently that the fail-safe mechanism has to be triggered. On the other hand, the average number of messages sent grows together with the percentage of relayer nodes until 70\% is reached, then with 80\% of relayers or more such a value plummets (see Figure~\ref{dandelion++-messages}).

\subsection{Gossip Algorithms Comparison}
\begin{figure}[h]%
    \centering
    \subfloat[\centering \label{coverage-progression} Number of messages spread in the graph when an algorithm that obtains a certain successful connection rate is used.]{{\includegraphics[width=.47\textwidth]{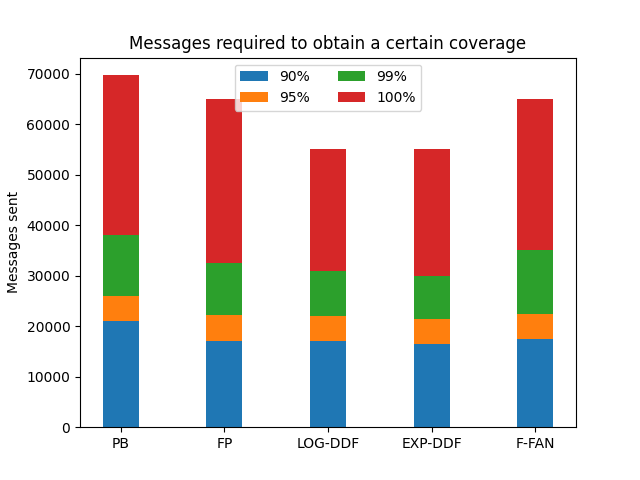} }}%
    \qquad
    \subfloat[\centering \label{delay} Delay overhead: cost in terms of time delivery for using a specific gossip algorithm instead of a pure broadcast scheme.]{{\includegraphics[width=.47\textwidth]{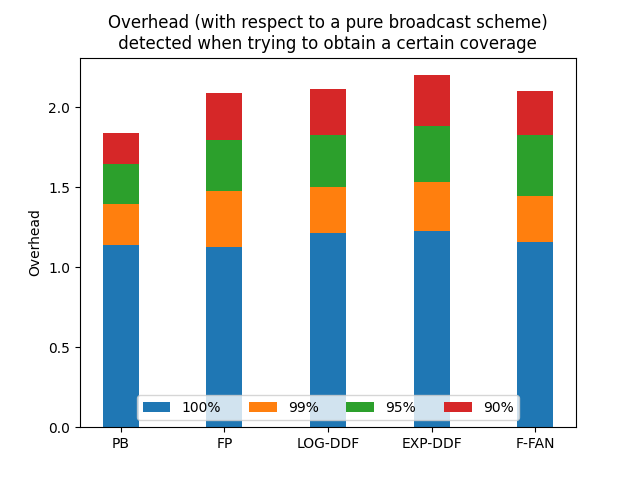} }}%
    \caption{Gossip protocols efficiency in relation with the pure broadcast scheme.}%
\end{figure}

The goal of using a probabilistic dissemination algorithm instead of a pure broadcast scheme (i.e.~relaying the messages to all the neighbors except the forwarder) is to minimize the network traffic. The cons of these protocols are that sometimes a minor coverage is achieved and a bigger amount of hops (and thus time) is necessary on average to deliver the messages. Not always minimizing the traffic networks entails sacrificing the coverage achieved. For example, in a graph with $8000$ active nodes and $15$ average connections per node, 100\% of successful communication rate is achieved with Fixed Probability having $60$ as parameter. That allows to save 40\% of the messages being sent with respect to when using a pure broadcast approach. The only cost in this case is the average delay, which is $1.12$ times bigger. The percentage of messages saved while still achieving a full coverage gets over 50\% when using a degree-dependent algorithm, this time with a $1.21$ times bigger average delay. However, degree dependent protocols have a downside, since they require from time to time to contact the neighbors to ask the number of peers they are connected to, thus generating an adjunctive network overhead.\\
\indent Figure~\ref{coverage-progression} and Table~\ref{tab-mess-saved} show that if 100\% of coverage is not necessary then the efficiency in terms of network traffic minimization can drastically increase. In fact, almost a half of the messages are necessary only to raise the successful communication rate from 99\% up to 100\%. Furthermore, experiments show that the degree dependent protocols are the best in terms of efficiency. However, the gap with the other algorithms is significant only when a successful communication rate over 95\% is needed. Probabilistic Broadcast, in addition, behave slightly worse than other protocols.
Figure~\ref{delay} shows the delay overhead comparison, pointing out that the number of hops necessary for delivering a message is inversely proportional to the efficiency of an algorithm in terms of traffic minimization. That is a consequence of the fact that, unlike when messages are broadcasted all the times, the communication rarely follows the shortest path connecting the sender and the recipient. Nevertheless for non-time-critical applications this downside is not particularly significant.

\begin{table}[h]
\centering
\caption{\% of messages w.r.t.~full broadcast, based on the achieved successful communication rate.}
\begin{tabular}{|c|c|c|c|c|}
\hline
- & \textbf{PB} & \textbf{FP}  & \textbf{DDF}  &   \textbf{F-FAN}\\ 
\hline
100 & 60.1\% & 56\% & 47.4\% & 56\% \\ 
99 & 32.8\% & 28\% & 26.4\% &  30.2\% \\
95 & 22.4\% & 19.2\% & 18.9\% & 19.4\%\\
90 & 18.1\% & 14.6\% & 14.7\% & 15.1\% \\
\hline
\end{tabular}
\label{tab-mess-saved}
\end{table}

\subsection{Hierarchical Networks}
\begin{figure}[h]
    \centering
    \begin{subfigure}[b]{0.495\textwidth}
    \includegraphics[width=\textwidth]{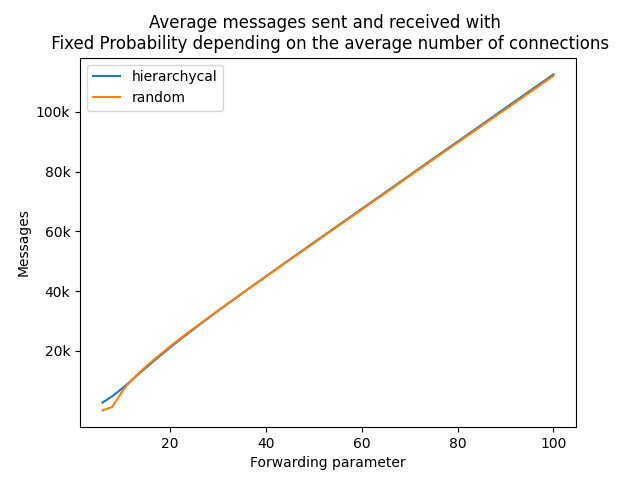}
    \end{subfigure}
    \begin{subfigure}[b]{0.495\textwidth}   
    \includegraphics[width=\textwidth]{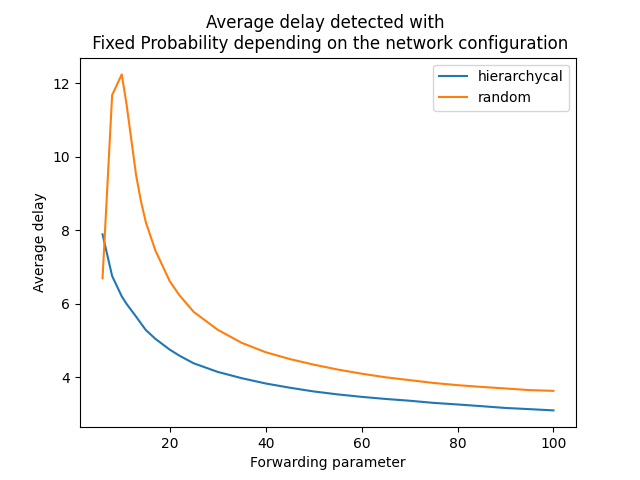}
    \end{subfigure}
    \caption{Effects of the average number of edges per node on the metrics.}
    \label{rand-scale}
\end{figure}

The previous results have been obtained using a random dynamic graph to represent the modelled network. In our experiments, the attachment of the nodes followed an aleatory procedure, meaning that the peers choose their neighbors randomly within the set of active nodes. Consequently, differences of degree among the nodes can happen but hubs are not generated.
However, it is possible to deal with networks in which there are some nodes with a considerably greater degree with respect to the rest of the nodes (i.e.~ hubs). So we have repeated the simulations using as a testbed a graph with a hierarchical structure, where a set of $80$ nodes (about 1\% of the active nodes) has a very high degree, between $250$-$300$ connections. In this new network configuration the total number of links is still around ${120\,000}$, entailing a consequent decrease of the average degree for the non-hubs nodes (moving from an average of $15$ connections to $12$).
While no considerable variation for the successful communication rate was noticed, Figure~\ref{rand-scale} shows that hierarchical networks ensure a lower delay without any significant variation on the number of messages sent. This is probably due to the presence of hubs, which reduce the average shortest path length among the nodes.

\section{Conclusions}
In this paper, we presented LUNES-Temporal, a new simulator for the modeling of dynamic networks, where connections among nodes have a temporal duration. While the model is general and several types of dynamic networks can be easily modeled using our tool, in this work we specifically focused on distributed and decentralized systems. The reason is that the overlying layer, built on top of the temporal network model, allows to simulate different dissemination strategies, some of which are used in certain DLTs (i.e.~Dandelion++).
Decentralized systems may require a significant traffic overhead, which is directly proportional to the average degree of the nodes. However, through accurate optimization, it is possible to reduce the traffic overhead. Our study shows that degree dependent function protocols are the most efficient algorithms to save as many messages as possible, having as downside a greater average delay to deliver the data. This kind of protocol, however, requires knowledge of the degree of the neighbors, so an extra communication overhead is needed. If this information is not available or is not easily retrievable, Fixed Probability and Fixed Fanout are still good alternatives.
In addition we have analyzed how the original proposal for Dandelion was troublesome, but such flaws can be avoided with the introduction of some features introduced by Dandelion++, like the fail-safe mechanism. The anonymity algorithms bring a greater delivery time, but this should not be an issue since they are proposed for blockchain environments, where a fast delivery of the messages is not the only concern.

\section*{Acknowledgments}
This work has received funding from the EU H2020 research and innovation programme under the MSCA ITN European Joint Doctorate grant agreement No 814177, LAST-JD-RIoE.

\bibliographystyle{scsproc}
\bibliography{biblio}

\section*{AUTHOR BIOGRAPHIES}

\noindent {\bf LUCA SERENA} is a research fellow at CIRI ICT, University of Bologna. His research interests include simulation, distributed systems, computer security and blockchain technologies.
His email address is \email{luca.serena2@unibo.it}. \\

\noindent {\bf MIRKO ZICHICHI} is a doctoral researcher in the Law, Science and Technology Joint Doctorate - Rights of Internet of Everything, funded by Marie Skłodowska-Curie Actions. His doctoral research focuses on the use of Distributed Ledger Technologies and Smart Contracts for the protection and distribution of individuals' personal data.
His email address is \email{mirko.zichichi@upm.es}. \\

\noindent {\bf GABRIELE D'ANGELO} is an Assistant Professor at the Department of Computer Science and Engineering, University of Bologna. His research interests include parallel and distributed simulation, distributed systems, online gaming and computer security. Since 2011 he has been in the editorial board of the Simulation Modelling Practice and Theory (SIMPAT) journal published by Elsevier.
His email address is \email{g.dangelo@unibo.it}. \\

\noindent {\bf STEFANO FERRETTI} is an Associate Professor at the Department of Pure and Applied Sciences, University of Urbino "Carlo Bo", since 2020. His research interests include distributed systems, complex networks, data science, fintech and blockchain technologies, multimedia communications, hybrid and distributed simulation.
His email address is \email{stefano.ferretti@uniurb.it}. \\

\end{document}